# Interface-driven electrical magnetochiral anisotropy in Pt/PtMnGa bilayers

Kangkang Meng*, Jikun Chen, Jun Miao, Xiaoguang Xu and Yong Jiang*

*School of Materials Science and Engineering, University of Science and Technology Beijing, Beijing 100083, China*

**Nonreciprocal charge transport, which is frequently termed as electrical magnetochiral anisotropy (EMCA) in chiral conductors, touches the most important elements of modern condensed matter physics. Here, we have investigated the EMCA in Pt/PtMnGa (PMG) bilayers with the assitance of nonequilibrium fluctuation theorems. Large EMCA in the Pt/PMG bilayers can be attributed to nonreciprocal response of an interface-driven chiral transport channel. Due to the presence of large charge fluctuations for small current region, higher order EMCA coefficients should be added and they are all functions of current. A combination of asymmetrical electron scattering and spin-dependent scattering furnish the PMG thickness dependent chiral transport behaviors in Pt/PMG bilayers. The dramatically enhanced anomalous Hall angle of PMG further demonstrates the modified surface state properties by strong spin-orbit coupling.**

*Corresponding authors: kkmeng@ustb.edu.cn; yjiang@ustb.edu.cn

Nonreciprocal response that excitations can only propagate in one direction of the materials system with broken inversion provides rich physics and functionalities [1-4]. Two well-known examples are the integer quantum Hall effect (QHE) that low-energy excitations can propagate only clockwise or counterclockwise, and the diode in semiconductor p-n junction [5]. Such nonreciprocal response can occur ubiquitously in various systems with noncentrosymmetry when the time-reversal symmetry is further broken by applying a magnetic field or with spontaneous magnetization, which is frequently termed as electrical magnetochiral anisotropy (EMCA) for electronic and magnetotransport in chiral conductors [6-14]. The EMCA has been phenomenologically discussed by Rikken, who provided a description for the directional transport in the noncentrosymmetric system by generalizing Onsager-Casimir reciprocal relations as follows [2]:

$$R(\boldsymbol{I},\boldsymbol{B}) = R_0(1+\beta \boldsymbol{B}^2 + \gamma \boldsymbol{I}\cdot\boldsymbol{B}) \quad (1)$$

where $R_0$, $\boldsymbol{I}$, and $\boldsymbol{B}$ represent the resistance at zero magnetic field, the electric current, and magnetic field, respectively. The parameter $\beta$ describes the normal magnetoresistance that is allowed in all conductors, and $\gamma$ is a coefficient tensor, representing the magnitude of the nonreciprocal resistance. Quite recently, from the viewpoint of not only fundamental physics but also applications, the nonreciprocal transport are investigated in various kinds of quantum materials, i.e., molecular conductor [6], chiral magnets [7,8], polar semiconductor [9], superconductor [10], noncentrosymmetric oxide interfaces [11], heavy metal/ferromagnet (HM/FM) bilayers [12], and nonmagnetic/magnetic topological insulator bilayers [13,14].

Generally, the nonequilibrium fluctuation theorems can provide the basis of the nonreciprocal transport beyond the Onsager-Casimir reciprocal relations [15,16]. The linear response theory founded on Onsager-Casimir reciprocal relations provides a powerful tool to describe a variety of physical systems [17-20]. However, such relations are justified only when the systems are close to thermodynamic equilibrium. In the regime of nonequilibrium (e.g., in nonreciprocal transport), the fluctuation theorem is proposed to play a more generalized role [21-26]. Although the nonequilibrium fluctuation theorems have been intensively used to explain the transport behaviors in mesoscopic systems, their applications to explain EMCA are far more difficult and not well explored as yet. Here, we have used heavy metal Pt/PtMnGa (PMG) heterostructures to investigate the EMCA with the assistance of nonequilibrium fluctuation theorems. Large EMCA has been found in the Pt/PMG films, which can be attributed to the nonreciprocal response in a chiral transport channel at the Pt/PMG interface due to the combination of strong spin-orbit coupling (SOC) of Pt and non-linear spin textures of PMG. In addition to $\gamma \boldsymbol{I}\cdot\boldsymbol{B}$, higher-order dependence terms should be taken into account in Eq. (1), and all the coefficients depend on the current amplitude due to charge fluctuations based on nonequilibrium fluctuation theorems. A combination of asymmetrical electron scattering and spin-dependent scattering furnishes the PMG thickness dependent nonreciprocal transport behaviors in the Pt/PMG films, exhibiting distinct angular dependences. Through anomalous Hall effect (AHE) measurements, we have found that both extrinsic skew scattering and intrinsic terms have been dramatically altered, which results in an enhanced anomalous Hall angle in the PMG capped with the Pt layer and further demonstrates the modified quantum geometrical nature at the Pt/PMG interface. Our result offers not only a new platform to investigate the nonreciprocal responses, but also an improved explanation of EMCA based on nonequilibrium fluctuation theorems.

High-quality PMG films were firstly grown on MgO (001) substrates at 400 $^{\circ}$C by co-sputtering Pt and MnGa targets in a high vacuum magnetron sputtering system with a base pressure of $<5 \times 10^{-6}$ Pa. The PMG thickness $d$ has been varied from 2 to 12 nm. Then, the films were annealed in situ at 500 $^{\circ}$C for 2 hours, and were left to cool down to room temperature in situ. Finally, heavy metal Pt films with different thickness $t$ from 1.5 to 7 nm were deposited onto the PMG films at room temperature. Besides, controlled samples $SiO_2$(2 nm)/PMG($d$ nm)/MgO(001) films were also prepared using the same procedure. Using standard photolithography techniques, we have fabricated Pt/PMG/MgO (001) and $SiO_2$/PMG/MgO (001) films into Hall bars in the size of 5μm×100μm. Hall bar devices were wire-bonded to the sample holder and installed in a physical property measurement system (PPMS, Quantum Design) for transport measurements with a temperature range of 5-400 K. We performed DC transport measurements using a Keithley 2400 current source and a Keithley 2182A nanovoltmeter. For AC transport measurements, a Keithley 6221 current source and two Stanford Research SR830 lock-in amplifiers were used.

Fig. 1(a) shows a cross-section transmission electron microscopy (TEM) micrograph from a 6-nm-thick PMG film deposited on MgO (001), which demonstrates explicitly the sharp interface and epitaxial growth of the (001)-oriented PMG on the MgO (001) substrate. The inset of Fig. 1(a) displays a wide-view scanning-TEM image of the same lamella, confirming the preparation of continuous films. Based on this interfacial region, selected area electron diffraction patterns of the films have been carried out, which further demonstrates the high crystal quality and cubic structure of the PMG layer as shown in Fig. 1(b) (A clearer figure without circle denotations is shown in Supplementary Note 1) [27]. In addition, energy dispersive x-ray spectroscopy demonstrates the stoichiometry of the film to be $Pt_{56}Mn_{22}Ga_{22}$. Fig. 1(c) shows the 2$\theta$-$\theta$ x-ray diffraction (XRD) pattern measured for a heterostructure of Pt (1.5 nm)/PMG (6 nm)/MgO (001). In addition to the (002) peak of MgO substrate, evident face-centered-cubic (FCC) lattice (002) peak of the PMG film in the spectrum can be observed. Distinct interference fringes are also observed, indicating a high crystal quality, perfect interface as well as smooth surface. Fig. 1(d) shows the 360° $\Phi$ scans of the (202) planes of MgO substrate and PMG film, which also exhibits the epitaxial growth of the (001)-oriented PMG on the MgO (001) substrate. By the way, the $\Phi$ scans of (111) planes have also been carried out. However, we have not found any clear (111) peaks. According to reciprocal space map of the PMG (204) reflection as shown in Fig. 1(e), the in-plane and out-of-plane lattice constants were calculated to be 0.416 nm and 0.419 nm, respectively. All the crystall structure measurements indicate that the PMG seems like a $B$2-type full Heusler alloy, which consists of four interpenetrating FCC sublattices [28]. However, the lattice we have calculated is surprisingly small, which is physically unreasonable! Therefore, the fabricated epitaxial PMG film can only be identified as a cubic lattice, which has a completely random distribution of all three atoms. Fig. 1(f) shows the schematic for longitudinal ($R_{2\omega}$) and transverse ($R_{2\omega}^H$) second harmonic magnetoresistance measurements with rotating **B** in *x-y* ($\alpha$), *z-y* ($\beta$) and *z-x* ($\theta$) planes.

The in-plane and out-of-plane magnetization measurements of the Pt (1.5 nm)/PMG (6 nm)/MgO (001) films were carried out by a superconducting quantum interference device. It should be noted that we have also carried out the same measurements for pure MgO substrates. After subtracting the MgO magnetic signals from the total results of Pt/PMG/MgO films, the magnetic properties of Pt/PMG are shown in Fig. 2. At 5 K, the out-of-plane saturation magnetization is ~189 kA/m, which is much larger than the in-plane case ~88 kA/m as shown in Fig. 2(a). The inset shows the enlarged **M-B** curves, which all exhibit hysteresis around zero magnetic fields. The saturation magnetic field is ~2.5 T and ~3.5 T for the in-plane and out-of-plane axes, respectively. The saturation magnetizations along the two axes show different temperature dependence but all decreases to zero at about 370 K as shown in Fig. 2(b), indicating a paramagnetic phase transition. The magnetization measurements indicate that the PMG film seems to have a ferrimagnetic state, which becomes long range ordered at large magnetic field [29-31]. If the applied magnetic field is perpendicular to the film plane, the net magnetization is larger than the in-plane case. Notably, we have not found spin-flop transition even when a large magnetic field of 14 T was applied along the two directions, indicating a strong antiferromagnetic exchange coupling between Mn atoms (Supplementary Note 1).

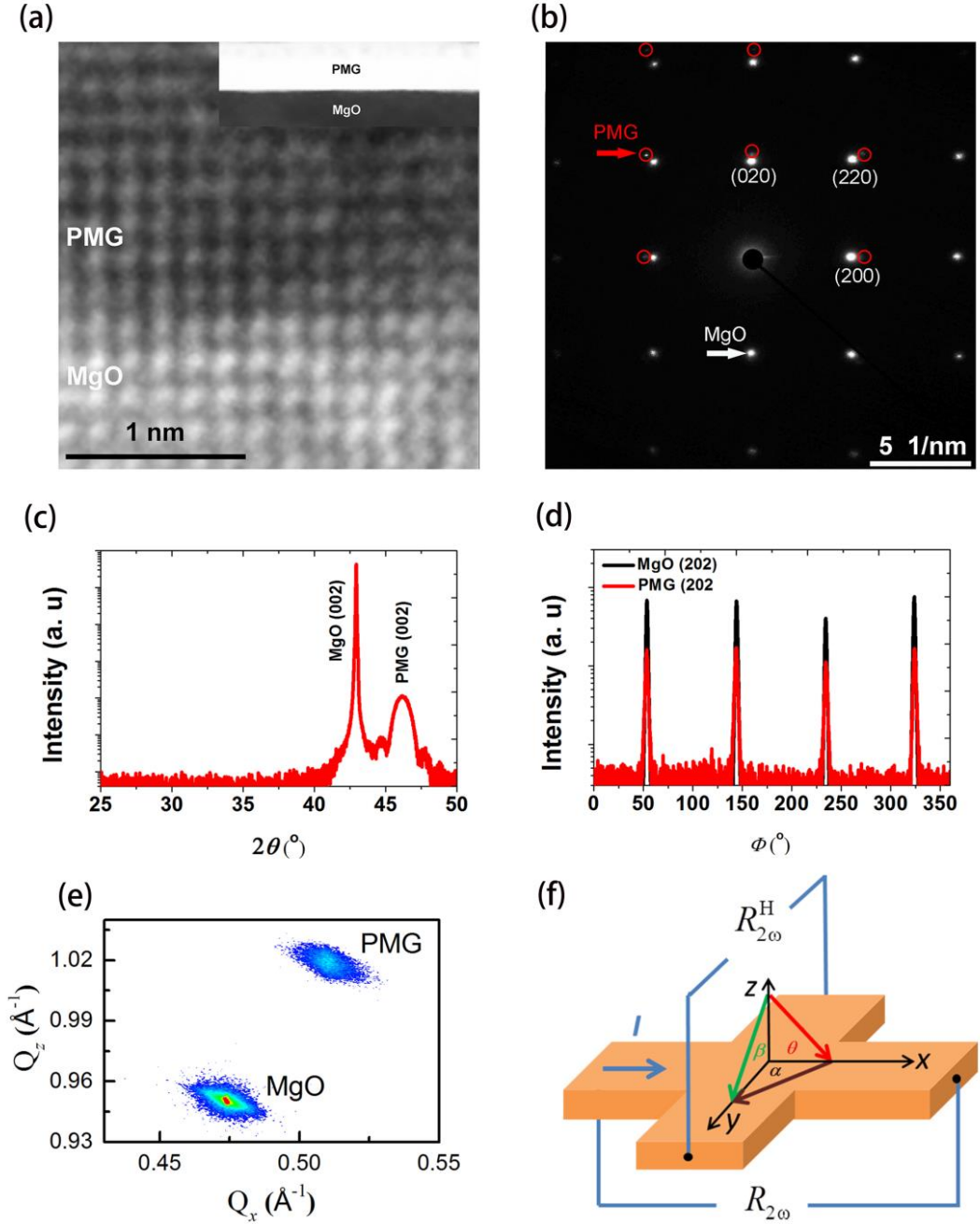

Fig. 1. (a) Cross-sectional TEM image of the multilayer thin films of Pt/PMG (6 nm)/MgO (001) films. (b) Selected area electron diffraction patterns of the interfacial region. (c) $2\theta$-$\theta$ XRD patterns measured for Pt (1.5 nm)/PMG (6 nm)/MgO (001) films. (d) 360° $\Phi$ scans of the (202) planes of MgO substrate and PMG film. (e) High-resolution XRD reciprocal space map of Pt (1.5 nm)/PMG (6 nm)/MgO (001) films. (f) Schematic for longitudinal ($R_{2\omega}$) and transverse ($R_{2\omega}^H$) second harmonic magnetoresistance measurements with rotating $B$ in $x$-$y$ ($\alpha$), $z$-$y$ ($\beta$) and $z$-$x$ ($\theta$) planes. The AC current is applied along $x$ direction.

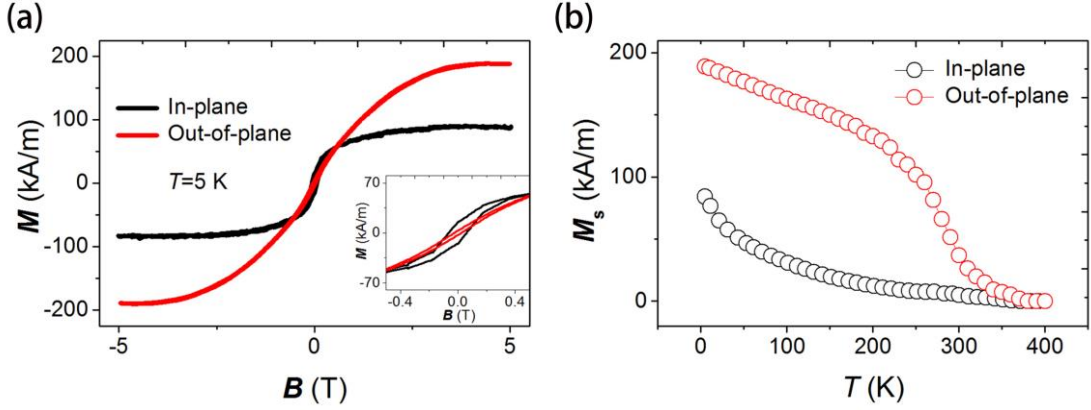

Fig. 2. (a) In-plane and out-of-plane *M-B* curves of Pt (1.5 nm)/PMG (6 nm)/MgO (001) films at 5 K. The inset shows the enlarged *M-B* curves at around zero magnetic field. (b) Temperature dependent saturation magnetization of Pt (1.5 nm)/PMG (6 nm)/MgO (001) films for the two directions.

Generally, the strong SOC of Pt can combine with the exchange interaction to generate an anisotropic exchange interaction that favors a chiral arrangement of the magnetization, which is known as Dzyaloshinskii-Moriya interaction (DMI) in the form of $\boldsymbol{D}_{12} \cdot (\boldsymbol{S}_1 \times \boldsymbol{S}_2)$ [32-34]. The vector $\boldsymbol{D}_{12}$ depends on the details of electron wave functions and could point to different directions, which depends on the symmetry and the precise crystalline structure. Contrary to the Heisenberg exchange interaction, which leads to collinear alignment of lattice spins, this form of DMI therefore very often cants the spins by a small angle [35-37]. Recently, using relativistic first-principles calculations, Belabbes et al. have shown that the chemical trend of the DMI in *3d-5d* ultrathin films follows Hund's first rule with a tendency similar to their magnetic moments in either the unsupported *3d* monolayers or *3d-5d* interfaces [35]. The largest absolute DMI value is obtained in Mn/5d films, indicating that DMI does depend critically not only on SOC and the lack of the inversion symmetry, but also on the *d* wave function hybridization of the *3d-5d* interface. Therefore, in addition to SOC, the Hund's exchange and crystal-field splitting of d orbitals could also induce large DMI at the Pt/PMG interface. Furthermore, the electrons can be scattered asymmetrically due to the emergence of $\boldsymbol{D}_{12} \cdot (\boldsymbol{S}_1 \times \boldsymbol{S}_2)$, resulting in the $\boldsymbol{S}_1 \times \boldsymbol{S}_2$ dependent EMCA [38]. Therefore, with spontaneous ferrimagnetism of PMG that breaks time-reversal symmetry, noncentrosymmetric interfaces combined with DMI induced by the Pt layer may bring out a similar chiral transport channel at the Pt/PMG interface [39]. We thus anticipate that EMCA could emerge at the Pt/PMG interface in the presence of magnetic field. To verify it, we have carried out the harmonic measurements with lock-in techniques. In our experiments, an AC current $I = I_0 \sin\omega t$ of frequency $\omega/2\pi$=125 Hz was applied, and the AC longitudinal ($V$) and transverse ($V^H$) voltages were recorded, respectively.

The longitudinal second harmonic resistance consists of three contributions, namely the EMCA ($R_{2\omega}^{EMCA}$), the magnetothermal effects due to the temperature gradients induced by Joule heating ($R_{2\omega}^{\nabla T}$), and the spin-orbit torque (SOT) induced modulation of the total magnetoresistance ($R_{2\omega}^{SOT}$):

$$R_{2\omega} = R_{2\omega}^{EMCA} + R_{2\omega}^{\nabla T} + R_{2\omega}^{SOT} \qquad (2)$$

Due to the symmetric behavior of the magnetoresistance with respect to the *x-y* plane, the out-of-plane oscillations driven by the damping-like SOT do not contribute to $R_{2\omega}$, while only Oersted and field-like effective fields contribute to $R_{2\omega}$. According to the work of Avci *et al.* [12], by taking into account the geometrical factor of the effective fields, the angular form of the SOT term is $R_{2\omega}^{SOT} \propto (\sin\alpha \cos^2\alpha)$.

Fig. 3(a) shows the in-plane (*x-y*) field-angle ($\alpha$) dependence of $R_{2\omega}$ for Pt (5 nm)/PMG (6 nm)/MgO (001) films with varying current from 1 to 5 mA and applying magnetic field 9 T at 5 K. The data shows a superposition of *n*th power of cosine function, and *n* equals to positive integer. As increasing current, the data shows more cosine-like field-angle dependence with a period of 360°. The $\sin\alpha \cos^2\alpha$ contributions are negligible for all current range, indicating that the SOT term has not contributed to $R_{2\omega}$. In addition, we have found that $R_{2\omega}^{\nabla T}$ is negligible (Supplementary Note 3). Therefore, it seems that the $\boldsymbol{I} \cdot \boldsymbol{B} = IB\cos\alpha$ term dominates the harmonic signals for larger current, indicating the emergence of EMCA in this film system. However, why high power terms become more evident for smaller current? According to the EMCA model of Rikken *et al.*, $\boldsymbol{I}$ and $\boldsymbol{B}$ should be integrally related to each other as $(\boldsymbol{I} \cdot \boldsymbol{B})$, and the $\gamma \boldsymbol{I} \cdot \boldsymbol{B}$ is a small term. We therefore infer that the two-terminal resistance of Pt/PMG films can be expressed as a polynomial of $(\boldsymbol{I} \cdot \boldsymbol{B})$:

$$R(\mathbf{I}\cdot\mathbf{B}) = R(0) + \gamma_1 \mathbf{I}\cdot\mathbf{B} + \frac{1}{2!}\gamma_2(\mathbf{I}\cdot\mathbf{B})^2 + \frac{1}{3!}\gamma_3(\mathbf{I}\cdot\mathbf{B})^3 + \cdots \quad (3)$$

where $R(0)$ represents the resistance that does not depend on $(\mathbf{I}\cdot\mathbf{B})$. For harmonic measurements, one has the longitudinal voltage as:

$$\begin{aligned}V(I) &= IR(\mathbf{I}\cdot\mathbf{B}) \\ &= R(0)I_0\sin\omega t + \gamma_1 I_0^2 B\cos\alpha\sin^2\omega t + \frac{1}{2!}\gamma_2 I_0^3 B^2\cos^2\alpha\sin^3\omega t + \frac{1}{3!}\gamma_3 I_0^4 B^3\cos^3\alpha\sin^4\omega t + \cdots \\ &= R(0)I_0\sin\omega t + \gamma_1 I_0^2 B\cos\alpha\sin^2\omega t + \frac{1}{2!}\gamma_2 I_0^3 B^2\cos^2\alpha\sin\omega t(1-\cos^2\omega t) + \frac{1}{3!}\gamma_3 I_0^4 B^3\cos^3\alpha\sin^2\omega t[1-\cos^2\omega t] + \cdots \\ &\approx [R(0) + \frac{1}{2!}\gamma_2 I_0^2 B^2\cos^2\alpha]I_0\sin\omega t + (\gamma_1 B\cos\alpha + \frac{1}{3!}\gamma_3 I_0^2 B^3\cos^3\alpha)I_0^2\sin^2\omega t + \ldots \\ &\approx [R(0) + \frac{1}{2!}\gamma_2 I_0^2 B^2\cos^2\alpha]I_0\sin\omega t + \frac{1}{2}(\gamma_1 B\cos\alpha + \frac{1}{3!}\gamma_3 I_0^2 B^3\cos^3\alpha)I_0^2[1+\sin(2\omega t - \pi/2)] + \ldots\end{aligned}$$
(4)

The phases of lock-in amplifiers were set to 0° and -90° for the first and second harmonic signal measurements, respectively. Then the first and second harmonic resistances are $\sim(R(0) + \frac{1}{2!}\gamma_2 I_0^2 B^2\cos^2\alpha)$ and $\sim\frac{1}{2}(\gamma_1 I_0 B\cos\alpha + \frac{1}{3!}\gamma_3 I_0^3 B^3\cos^3\alpha)$, respectively. Considering $R(0) \gg \frac{1}{2!}\gamma_2 I_0^2 B^2\cos^2\alpha$, the first harmonic resistance only reveals common anisotropic magnetoresistance (Supplementary Note 4). It should be noted that one can add much higher power terms into the second harmonic resistance, but only odd terms contribute according to our calculation. Then the data can be fitted well if the $R_{2\omega}$ was expressed as a superposition of odd power of cosine function. However, this conjecture cannot fully explain the phenomena since higher power terms become more important for smaller current (Supplementary Note 5). Therefore, we should note that all the coefficients $\gamma_i (i=1,2,3,...)$ should be functions of current. It indicates that the EMCA should be influnced by charge fluctuations in the chiral transport channel at the Pt/PMG interface based on nonequilibrium fluctuation theorems. Fluctuation theorem is an important theory that holds even in the far-from-equilibrium regime [21-26]. It can give the relation between the nonreciprocal transport coefficients and current noise in quantum transport of a mesoscopic conductor. The current *I* passing through a conductor can be expressed as a polynomial of the bias voltage *V* as:

$$I = G_1 V + \frac{1}{2!}G_2 V^2 + \frac{1}{3!}G_3 V^3 + \cdots \quad (5)$$

where the first term represents Ohm's law with conductance $G_1$, which is directly related to the transmission in a mesoscopic conductor. In contrast, the higher order coefficients ($G_2, G_3,...$) convey information on electron-electron interactions in a voltage biased conductor.

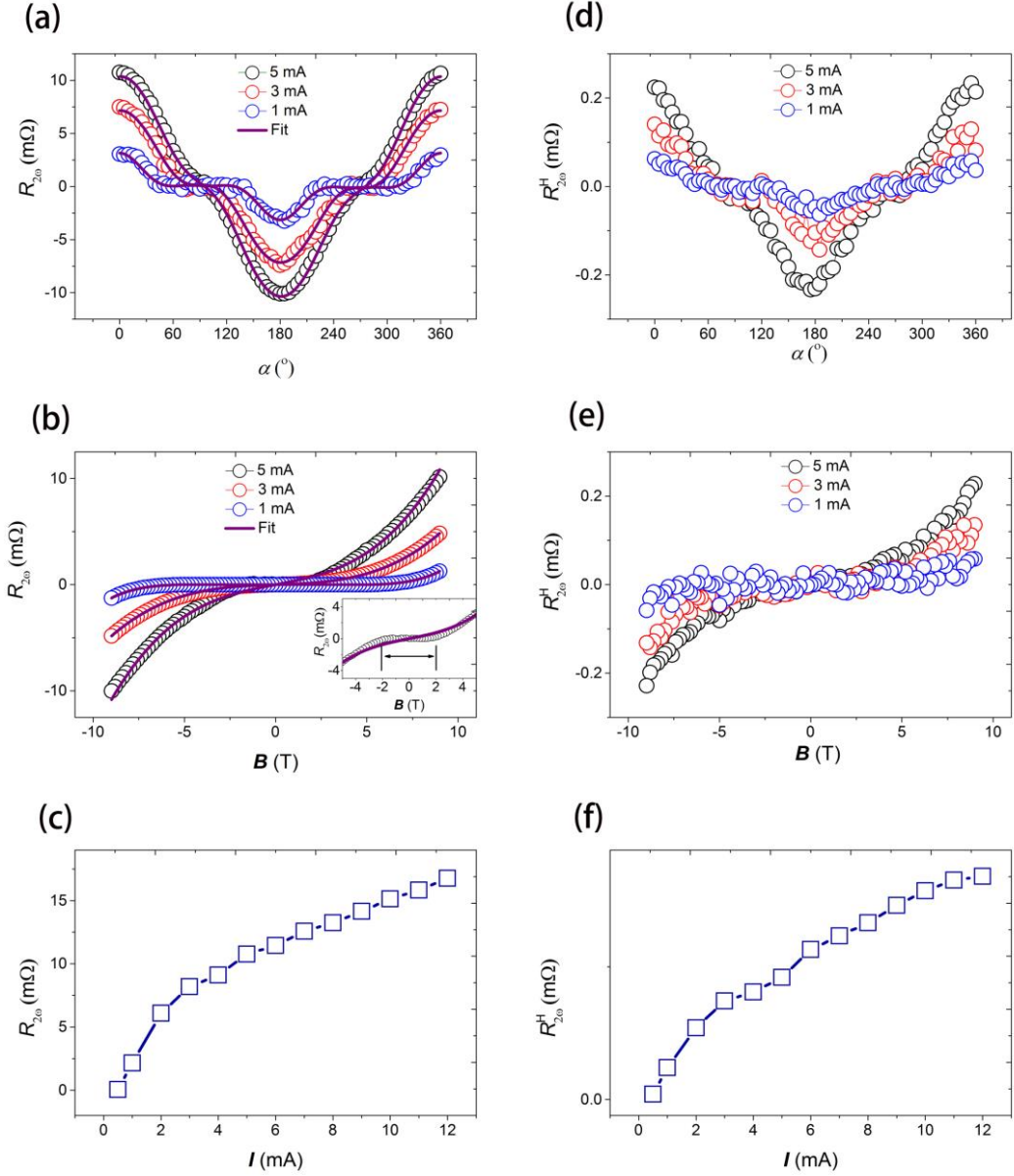

Fig. 3. (a) Angle ($\alpha$) dependence of the longitudinal second harmonic resistance $R_{2\omega}$ with varying applied AC current in Pt (5 nm)/PMG (6 nm)/MgO (001) films under 9 T at 5 K. Solid lines are fits to the experimental data according to Eq. (4). (b) Magnetic field dependence of $R_{2\omega}$ with varying applied AC current at 5 K. Solid lines are fits to the experimental data according to Eq. (4). The inset shows the enlarged curves with 5 mA at around zero magnetic field. (c) AC current amplitude dependence

of $R_{2\omega}$ under 9 T at 5 K. (d) Angle ($\alpha$) dependence of the transverse second harmonic resistance $R_{2\omega}^H$ with varying applied current under 9 T at 5 K. (e) Magnetic field dependence of $R_{2\omega}^H$ with varying applied AC current at 5 K. (f) AC current amplitude dependence of $R_{2\omega}^H$ under 9 T at 5 K.

Based on this conjecture, the $R_{2\omega}$-$B$ data with sweeping magnetic field parallel to current directions ($x$) at 5 K can also be fitted well as shown in Fig. 3(b). It should be noted that, in low field range (-2 T<$B$<2 T), $R_{2\omega}$ are almost zero, deviating from the fitted lines. As mentioned above, we have speculated a $D_{12} \cdot (S_1 \times S_2)$-indcued EMCA at the Pt/PMG interface due to the strong SOC of Pt and ferrimagnetic PMG film. However, it should be noted the formation of long range ordered non-coplanar magnetic configurations ($S_1 \times S_2$) may need large magnetic field according to the in-plane $M$-$B$ curves as shown in Fig. 2(a). Therefore, evident EMCA in the Pt/PMG films (non-zero $R_{2\omega}$) needs applying a large magnetic field and a finite magnetization (Supplementary Note 6). A similar non-monotonic magnetic field dependence of $R_{2\omega}$ has also been observed in chiral magnets MnSi and CrNb$_3$S$_6$ [7,8], and the current dependence of $R_{2\omega}$ showed linear behaviors, which were ascribed to chiral spin fluctuations at specific temperature-magnetic field-pressure regions. On the other hand, as mentioned above, the spin correlation with vector spin chirality in chiral magnets scatters electrons asymmetrically, resulting in the EMCA [38]. A similar scattering picture could also be depicted at the Pt/PMG interfaces, but the charge-fluctuations-induced non-linear $I$-$V$ characteristics needs consideration since we have found a non-monotonic current dependence of $R_{2\omega}$ in the Pt/PMG films at 5 K as shown in Fig. 3(c). On the other hand, considering the EMCA is an interface-driven phenomenon, the increased current shunting effect for the thicker PMG films could quench the EMCA (Supplementary Note 7).

Furthermore, nonlinear planar Hall effect is another proof to verify the emergence of nonreciprocal charge transport, though different determined mechanisms have been reported and remain elusive [14,40]. Fig. 3(d) shows the in-plane ($x$-$y$) field-angle ($\alpha$) dependence of second transverse harmonic magnetoresistance $R_{2\omega}^{H}$ for Pt (5 nm)/PMG (6 nm)/MgO (001) films with varying current from 1 to 5 mA and applying magnetic field 9 T at 5 K. Similar with second longitudinal harmonic magnetoresistance, the nonreciprocal transport coefficients of $R_{2\omega}^{H}$ also depend on current amplitudes and higher-order terms become more crucial for smaller current. Meanwhile, it can also prove that the SOT contribution in Pt (5 nm)/PMG (6 nm)/MgO (001) films is negligible since SOT will be quenched with applying large magnetic fields. On the contrary, $R_{2\omega}^{H}$ with varying current shows non-monotonic magnetic field dependence at 5 K as shown in Fig. 3(e), similar with $R_{2\omega}$-$B$ relationships. On the other hand, Fig. 3(f) shows that $R_{2\omega}^{H}$ also has non-monotonic relationship with current at 5 K. Therefore, neither asymmetric magnon-mediated scattering nor concerted actions of spin-momentum locking and time-reversal symmetry breaking is applicable to the nonlinear planar Hall effect in Pt (5 nm)/PMG (6 nm)/MgO (001) films [13,14,41], and asymmetrical electron scattering as well as charge fluctuations should determine the behaviors.

In order to gain further insight into the EMCA of the Pt/PMG films, we have investigated the AC transport properties with varying PMG thickness. Fig. 4(a) shows the in-plane ($x$-$y$) field-angle ($\alpha$) dependence of $R_{2\omega}$ for Pt (5 nm)/PMG (2 nm)/MgO (001) films with varying current from 1 to 5 mA and applying magnetic field 9 T at 5 K. Contribution from unidirectional magnetoresistance (UMR) in HM/FM bilayers ($R_{2\omega}^{UMR}$) should be added into in Eq. (4) [12]:

$$R_{2\omega} = R_{2\omega}^{EMCA} + R_{2\omega}^{UMR} + R_{2\omega}^{\nabla T} + R_{2\omega}^{SOT}. \quad (6)$$

where $R_{2\omega}^{UMR} = C'(I)IB\sin\alpha$, and the coefficient $C'(I)$ should also be a function of current based on fluctuation theorems as discussed above. Both $R_{2\omega}^{\nabla T}$ and $R_{2\omega}^{SOT}$ can also be neglected, which have been discussed in Supplementary Note 4 and the following paragraphs, respectively. Here, to concisely elucidate the current dependent coefficient tensor $\gamma_i$, we have also neglected much higher oreder terms ($i \geq 6$). On the other hand, we have just reserved the first power term of UMR for simple treatment. The emergence of UMR can be ascribed to the decrease of the PMG thickness, since UMR will become evident if the thickness of the FM layer is comparable to the spin diffusion length [42]. When the FM thickness is much larger than the spin diffusion length, more current is shunted into the bulk of the layers. Therefore, the UMR induced by the interface spin-dependent scattering gets diluted in Pt(5 nm)/PMG(6 nm)/MgO (001) films.

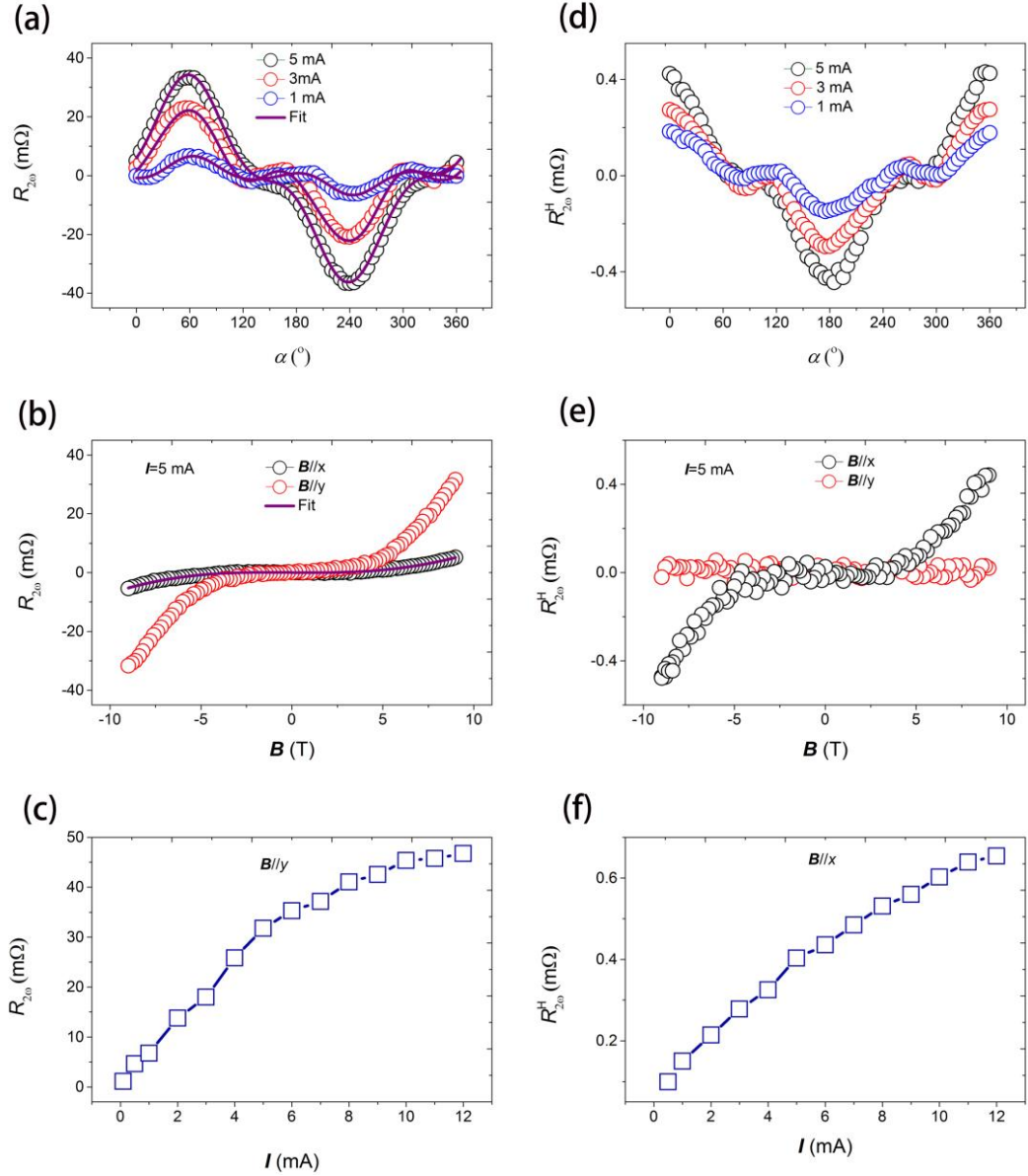

Fig. 4. (a) Angle ($\alpha$) dependence of the longitudinal second harmonic resistance $R_{2\omega}$ with varying applied current in Pt (5 nm)/PMG (2 nm)/MgO (001) films under 9 T at 5 K. Solid lines are fits to the experimental data according to Eq. (6). (b) Magnetic field ($B//x$ and $B//y$) dependence of $R_{2\omega}$ with an applying AC current 5 mA at 5 K. Solid lines are fits to the experimental data according to Eq. (6). (c) AC current amplitude dependence of $R_{2\omega}$ with applying magnetic field 9 T along $y$ direction at 5 K. (d) Angle ($\alpha$) dependence of the transverse second harmonic resistance $R_{2\omega}^H$ with varying applied AC current. (e) Magnetic field ($B//x$ and $B//y$) dependence of $R_{2\omega}^H$

with $I$=5 mA. (f) AC current amplitude dependence of $R_{2\omega}^H$ with applying magnetic field 9 T along $x$ direction at 5 K.

Generally, there are three competing mechanisms underpinning the resistance asymmetry, namely, interface and bulk spin-dependent electron scattering and electron-magnon scattering. The first and second mechanisms have been referred as the interface and bulk spin-dependent (SD) UMR, respectively, and the third as spin-flip (SF) UMR [42]. As shown in Fig. 4(b), $R_{2\omega}$ becomes larger with the increasing magnetic field along $y$ direction at 5 K, so the SF-UMR contribution is negligible since the SF-UMR signals should become small due to the quenching of electron-magnon scattering at high magnetic fields. Furthermore, $R_{2\omega}(\boldsymbol{B})$ along $x$ direction can also be fitted based on Eq. (4). We should note that, according to the previous report of the UMR in HM/FM bilayers, $R_{2\omega}$ should become a constant dominated by SD-UMR, in analogy with the current-in-plane giant magnetoresistance [12,42]. Therefore, it cannot fully explain the non-monotonic magnetic field dependence along $y$ direction as shown in Fig. 4(b). One possible reason is the magetic textures of PMG. Though in-plane saturation magnetization has been found at ~2.5 T as shown in Fig. 2(a), the interface magetic textures could be affected by large magnetic field, modulating interface scattering potential induced by the spin Hall effect and the ensuing interfacial resistance change. Another possible reason is the strong Rashba-type SOC due to the enhanced broken of space inversion symmetry when the PMG thickness is 2 nm. Choe *et al.* have found that the relative strength of the Rashba SOC in comparison with the Fermi energy is a governing factor to produce the nonreciprocal charge transport in the conductive oxide interface of LaAlO$_3$/SrTiO$_3$ [43]. This is conceivable, as the nonreciprocal charge transport is in principle a perturbation effect by SOC. Ideue *et al*. have identified that Fermi energy and Rashba splitting should determine the nonreciprocal charge transport in a three-dimensional Rashba-type polar semiconductor BiTeBr [9]. Considering the giant Rashba SOC, the appearance of polynomial growth of $R_{2\omega}(\boldsymbol{B})$ would be associated with a relative strength between magnetic field and Rashba SOC. Furthermore, it should be noted that, though the thickness of PMG is rather small, which is comparable with the spin diffusion length leading to the UMR discussed above, SOT is also negligible in Pt (5 nm)/PMG (2 nm)/MgO (001) films. Fig. 4(d) shows the in-plane (*x-y*) field-angle ($\alpha$) dependence of transverse second harmonic magnetoresistance $R_{2\omega}^H$ for Pt (5 nm)/PMG (2 nm)/MgO (001) films with varying current from 1 to 5 mA and applying magnetic field 9 T at 5 K. The larger $R_{2\omega}^H$ with the increasing magnetic field shown in Fig. 4(e) cannot be ascribed to SOT, indicating a strong quench of spin torques at the interface, though a large spin current could be injected from Pt. It further demonstrates that the EMCA in the Pt/PMG films has no relation with the spin current transport in bulk PMG. By the way, the fluctuation theorems also determine the non-linear $R_{2\omega}$-*I* and $R_{2\omega}^H$-*I* relationships at 5 K as shown in Figs. 4(c) and (f), respectively. Furthermore, out-of-plane scan of first and second longitudinal harmonic resistance in Pt (5 nm)/PMG (*d* nm)/MgO (001) films and other contributions to nonreciprocal transport have been discussed in Supplementary Note 8.

Microscopically, in chiral transport systems, the nonreciprocal transport phenomena are frequently encoded by the Berry phase and the asymmetric scattering [4], which can be analyzed through discussing the AHE contributions [44]. Therefore, we have further investigated the AHE in Pt (*t* nm)/PMG (*d* nm)/MgO (001) films. By subtracting the ordinary Hall term from the total Hall resistivity, we have obtained the anomalous Hall resistivity $\rho_H$ of all the multilayers under perpendicularly applied magnetic field ***B*** in the temperature range from 5 to 400 K (Supplementary Note 9). It should be noted that, assuming that each film in the multilayers acts as a parallel resistance path [34], both anomalous Hall resistivity $\rho_H$ and longitudinal resistivity $\rho_{xx}$ have been expressed as those of the PMG layers. Fig. 5(a) shows the representative relationships between $\rho_H$ and the square of longitudinal resistivity $\rho_{xx}^2$ in Pt (*t* nm)/PMG (6 nm)/MgO films, in which the sample of *t*=0 nm refers to SiO$_2$ (2 nm)/PMG (6 nm)/MgO films. It is found that the experimental data can be described by scaling law $\rho_H = a\rho_{xx0} + b\rho_{xx}^2$, where $\rho_{xx0}$ is the residual resistivity induced by impurity scattering, *a* and *b* denote extrinsic skew scattering and intrinsic mechanism, respectively [45]. We have found that $\rho_H$ is significantly increased after capping Pt layers, indicating an enhanced anomalous Hall angle. The Pt-thickness dependent *a* and *b* in Pt (*t* nm)/PMG (6 nm)/MgO films is shown in Fig.

5(b). Both *a* and absolute value of *b* increase sharply when *t* is varied from 0 to 1.5 nm, and then the values keep almost constant with further increasing *t*. It indicates that both skew scattering and intrinsic mechanism related to the Berry phase contributions have been enhanced after capping Pt layers, which should be determined by the strong interface SOC introduced by Pt. This behavior has also been found in Pt (*t* nm)/PMG (2 nm)/MgO (001) films as shown in Figs. 5(c) and (d), but the enhancement of $\rho_H$ is much more significantly especially for low temperature regions as compared with Pt (*t* nm)/PMG (6 nm)/MgO (001) films. Therefore, the Pt layer has dramatically modified the surface state properties of PMG. Combinated with ferrimagnetic PMG, the strong SOC may induce a chiral transport system at the interface.

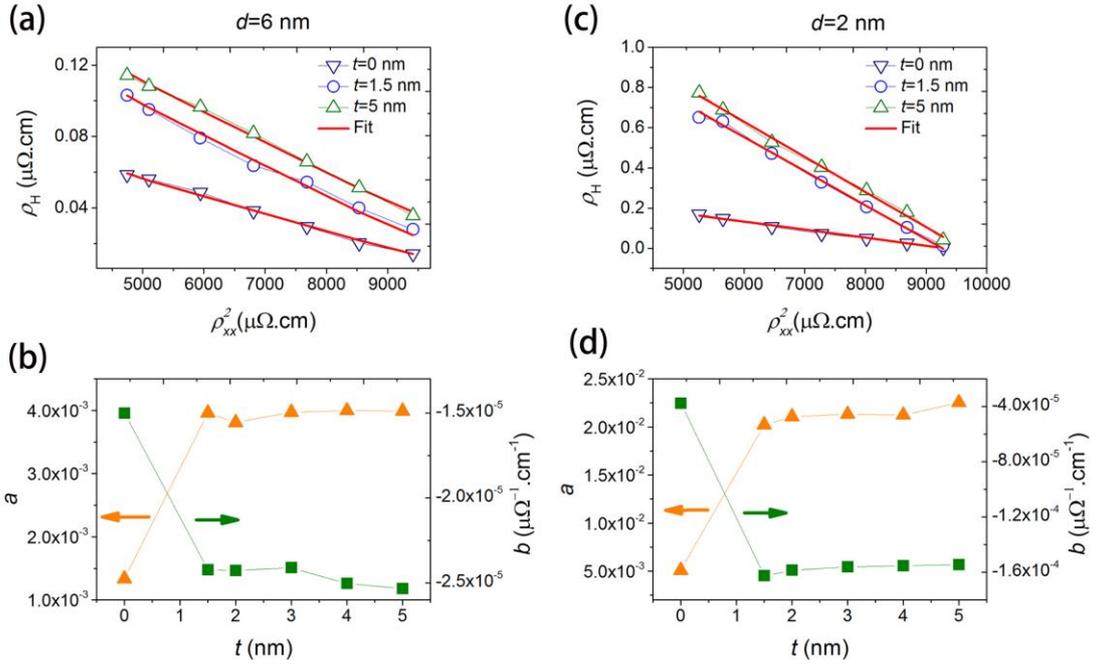

Fig. 5. (a) $\rho_H$ - $\rho_{xx}^2$ in Pt (*t* nm)/PMG (6 nm)/MgO (001) films with different *t*. The sample of *t*=0 nm refers to SiO$_2$ (2 nm)/PMG (6 nm)/MgO (001) films. Solid lines are fits to the experimental data according to scaling law $\rho_H = a\rho_{xx0} + b\rho_{xx}^2$. (b) Pt thickness (*t*) dependence of coefficients *a* and *b* in Pt (*t* nm)/PMG (6 nm)/MgO (001) films. (c) $\rho_H$ - $\rho_{xx}^2$ in Pt (*t* nm)/PMG (2 nm)/MgO (001) films with different *t*. The sample of *t*=0 nm refers to SiO$_2$ (2 nm)/PMG (2 nm)/MgO (001) films. Solid lines are fits to the experimental data according to above scaling law. (d) Pt thickness (*t*) dependence of coefficients *a* and *b* in Pt (*t* nm)/PMG (2 nm)/MgO (001) films.

Finally, we want to mention that: (i) It is hard to give an accurate explaination that why the charge fluctuation or current noise was so large in the low current region of Pt/PMG films. The length of our Hall device is ~100 μm, which should be much larger than the mean free path. Therefore, the mesoscopic view of electronic motion is that it should be "diffusive" [46,47]. However, in the metallic diffusive regime, the effects of electron-electron interactions may be very important after depositing Pt layer, which can increase the current noise. The transport channel at the Pt/PMG interface is more like a quasi-two-dimensional conductor, and the current noise could depend on transmission coeffcients, the weak localization, the disordered interfaces and so on. (ii) We cannot give a definite magnetic texture of PMG at the Pt/PMG interface, and we can only speculate a chiral transport channel through the unusual transport behavior of Pt/PMG. Additional precise magnetic experiments and theoretical work are needed.

In conclusion, we have investigated the EMCA of the Pt/PMG films and improved the framework of EMCA based on nonequilibrium fluctuation theorems. In the presence of large magnetic field, the coefficients of EMCA in DMI-induced chiral transport channels are functions of current. With varying the PMG thickness, the EMCA in the Pt/PMG films exhibits distinct nonreciprocal transport behavior which is ascribed to the competition of asymmetrical electron scattering and spin-dependent scattering. The modified Berry phase and scattering properties due to the strong SOC of Pt at the interface have also been demonstrated through the AHE measurement. Our results suggest this improved explaination of EMCA based on nonequilibrium fluctuation theorems may be applicable to the chiral transport properties in a variety of quantum materials.

**ACKNOWLEDGMENTS**

This work was partially supported by the National Key Research and Development Program of China (2019YFB2005801), the National Natural Science Foundation of China (Grants No. 51971027, No. 51731003, No. 51971023, No. 51927802, No. 51971024), and the Fundamental Research Funds for the Central Universities (Grants No. FRF-TP-19-001A3).